# Thermodynamics deposition on curve nanoholes

O. A. Pinto[1], B. A. López de Mishima[1], E. P. M. Leiva[2], and O. A. Oviedo[2]

1) Centro de Investigaciones y Transferencia de Santiago del Estero, (CITSE-CONICET) Universidad de Santiago de Estero, RN 9 Km 1125 Villa el Zanjón, Santiago del Estero, CP 4206 Argentina.

2) Instituto de Fisicoquímica de Córdoba, Departamento de Matemática y Física de la Facultad de Ciencias Químicas, Universidad Nacional de Córdoba, Córdoba, Argentina. X5000HUA.

**Abstract**

In this work the thermodynamics of the electrodeposition on nanoholes is analyzed. Different lattice-gas models of nanoholes, from parallelepiped geometry to the empty bulk of a nanoparticle, were considered. The models include curvature on the inner walls. Several stages of deposition are identified. Monte Carlo technique in the Grand Canonical Ensemble is used to determine isotherms, isosteric heat, energy per site, etc. The study is based on different ranges of energies and nanoholes sizes.

## 1. Introduction

In the last years new experimental techniques allow building new structures on a small scale with amazing properties. In a small scale, the position of the atoms in the structure is critical for their contribution to the thermodynamics properties. The thermodynamics properties in nanostructures depend on the different coordination of the constituent atoms. In particular, size and shape affect the physical-chemical properties of the small systems, for example, in: melting of small particles [1,2], chemical reactivity [3], diffusivity [4], sintering [5], and under/over potential deposition transition [6,7], etc. In a small scale, the traditional thermodynamics functions have been modified in several ways. A visionary theoretical work was done by T. Hill[8,9] in the 1960s. This work describes the influence of a few atoms in the global thermodynamics properties, and introduces a new name for this area, nanothermodynamics.

Several experimental techniques were used to study and generate this kind of nanosystems. In particular, by using STM technique it is possible to generate defects that can be a center of growth for metal depositions. A review of this application is discussed in [10]. This technique can also be used to generate pinholes, nanocavities or nanoholes in metal surfaces. Nanoholes (NH) were created on a graphite surface, and then filled with Ag from the STM tip [11]. In Au(111) surfaces, NH were created by short negative-voltage pulses to the STM tip and then filled with Cu clusters[12]. The NH depth was of three monolayers. The nucleation sites are provided by the edges or steps in the NHs for the Cu. Other metals analyzed were Bi and Ag on Au(111) with the same technique as before[13].

The surface decoration of nanoparticles with different geometries has been studied with Monte Carlo simulations. Nanoeffects, such asa shift of adsorption isotherms with the nanoparticle size, were reported in [14]. It is important to mention that in 2D systems the isotherms have no significant shifts when the size of the substrates changes. Theoretical approximations like the detailed mean-field have been done in [15]. The comparison with simulations shows a good concordance. Systems such as Cu/Au(111) and Ag/Au(111) were studied with off-lattice models and embedded atom method (EAM) [16]. In Cu/Au(111) sites linked with four atoms on the inner walls as well as nucleation sites and cluster formations were identified. But in Ag/Au(111) the filling of the nanoholes was described by a layer-by-layer mechanisms.

The main objective of this work is the characterization of the thermodynamics deposition of NH. Different geometries were considered. The wall curvature is taken as a point of comparison. Lattice gas model and the technique of the electro-deposition were used. For this purpose we use Grand Canonical Monte Carlo method, and we set two kinds of energies: intra and inter-particle.

**2. Models and basic definitions**

The solid substrate is tackled with a lattice gas model. All foreign particles can be deposited in a specific adsorption site in the solid. The deposition process is controlled by electrochemical potential. Three models were considered. Model A represents a tridimensional solid where the NH has parallelepiped geometry. The solid is formed by four planes of atoms separated by a lattice constant. On three of them a hole is created. The walls are vertical with no curvature. On the external surface, atoms can be deposited and form a monolayer. In this case whole solid has {100} geometry. The red dots represent the atoms in the solid and the white dots represent the empty sites. The surface sites can be written as $(L+l)^2$, where $L$ is the lateral size of the hole and $l$ is the constant distance between edges of NH and solid. Each plane has $L^2$ sites. Then the volume is $V_{NH-A} = 3L^2$.

Model B mimics a curvature on the walls. As in the previous model, four planes are considered. The curvature is regarded as including a step per each plane. Empty sites are $L^2$, $(L-2)^2$, $(L-4)^2$ for planes 3, 2 and 1-respectively. The volume is $V_{NH-B} = L^2 + (L-2)^2 + (L-4)^2$. In Model C, other case of curvature is considered. The deposition sites of NHs are the bulk sites in half of NP. Anicosahedra was considered. In this model the coordination of the inner sites changes dramatically. The center of the bulk of the icosahedra has twelve nearest neighbors (NN), but this number decreases when the sites are near the shell. The monolayer is not considered. Snapshot of the models are shown in figure 1 for L=30, l=5. The insets a), b) and c) correspond to models A, B and C respectively. The blue lines are guide eyes.

The solid substrate is formed by particles called α. Meanwhile, the adsorbates are β. The last particles are characterized by electrochemical potential $\mu$ at temperature $T$. Only nearest neighbors are considered. From the perspective of the interactions, the adsorbates can "feel" two kinds of energies: the interparticles $w$, and intraparticles ε. The Hamiltonian can be written as:

$$H = \sum_{\langle (i,j,k),(i',j',k') \rangle} c_{i,j,k} c_{i',j',k'} \left[ \left( \frac{c_{i',j',k'} - 1}{2} \right) w + \left( \frac{c_{i',j',k'} + 1}{2} \right) \varepsilon \right] \quad (1)$$

The indices (i,j,k) correspond to the spatial coordinates of the solid. $c_{i,j,k}$ is zero if the site (i,j,k) is empty, 1 if occupied by a β-particle. The summation goes to the adsorption sites only. Multiple occupations and the tunnel effect are forbidden.

The deposition process is simulated by a standard importance sampling Monte Carlo method in the Grand Canonical Ensemble [17, 18, 19]. To satisfy the principle of detailed balance, the Metropolis algorithm [20] was used. A Monte Carlo Step (MCS) corresponds to M attempts to change the state of the system. Before sampling the quantities of interest, thermodynamic equilibrium was established. For all simulations, the equilibrium state was obtained discarding the first $5 \times 10^6$ MCSs. Then, the next $2 \times 10^6$ MCSs were used to compute averages. The parameters of interest can be obtained by simple averages.

The coverage of individual planes, or partial coverage, is a very useful parameter for the thermodynamics analysis. The partial coverage is:

$$\theta_i(\mu) = \frac{\langle N_{i,\beta} \rangle}{L^2} \quad i=1,2,3 \quad \text{Model A}$$

$$\theta_1(\mu) = \frac{\langle N_{1,\beta} \rangle}{L^2}, \quad \theta_2(\mu) = \frac{\langle N_{2,\beta} \rangle}{(L-2)^2}, \quad \theta_3(\mu) = \frac{\langle N_{3,\beta} \rangle}{(L-4)^2} \quad \text{Model B} \quad (2)$$

The indices "$i$" indicate each plane, then $N_{i,\beta}$ is the quantity of particles deposited in each $i$-plane. If $i=1$ is the bottom plane, $i=2$ is the middle plane, finally $i=3$ is the higher plane. The full coverage, then, is computed as:

$$\theta(\mu) = \sum_i \theta_i(\mu) \quad (3a)$$

Or

$$\theta(\mu) = \frac{1}{V_{NH}} \sum_{(i,j,k)} \langle c_{i,j,k} \rangle \quad (3b)$$

As the total isotherm is normalized to volume, when $\theta=1$, the NH is full. The energy per volume is:

$$u = \frac{\langle H \rangle - \mu \langle N_\beta \rangle}{V_{NH}} \quad (4)$$

The differential heat of deposition is:

$$q_d = -\frac{\partial u}{\partial \theta} = \frac{\langle N_\beta H \rangle - \langle H \rangle \langle N_\beta \rangle}{\langle N_\beta^2 \rangle - \langle N_\beta \rangle^2} \quad . \quad (5)$$

In all cases $<\cdots>$ means the average over the MC simulation run. For Model C, the equation (3b), (4) and (5) can be applied, except that $V_{NH}$ are sites in the bulk of the NH.

## 3. Discussion

In this section the thermodynamics of the three models are described.

**3.1 Model A**

In this model the energies considered were: $w/k_BT$=-1.0 and $\varepsilon/k_BT$ <0.0. The attractive energies were chosen because they try to mimic the metal-metal interactions. $\varepsilon/k_BT$ takes into account interactions between different metals. The energies were settled as: $\varepsilon/k_BT$ =0.0, -0.5, 0.75, -1.0, -2.0,-5.0 and -7.0. The size of the NH was L=30. Here, the adsorbate tries to adsorb in sites where coordination is maximum.

Figure 2 shows the adsorption isotherms, and it should be noticed that the isotherms do not saturate at the same point. This occurs because there are sites that cannot be filled, which is a consequence of the prohibition of the tunneling of the adparticles. When a particle is deposited, for example on plane 3 next to the walls, the empty sites below cannot be filled until the first particle is removed. But when $\varepsilon/k_BT$ is more negative, the β particles prefer to link with α particles, and the vacancies are gradually removed.

The maximum coverage can be calculated by $\theta_{max}(L,l) = 1 + (L+2l)^2/3L^2$, then $\theta_{max}(L=30, l=5) = 43/27$. For $\varepsilon/k_BT$<-5, all isotherms saturate at the same value. When $w/k_BT = \varepsilon/k_BT$ =-1.0, the adsorbate has the same probability of sticking on NH and surface sites. For $\varepsilon/k_BT > w/k_BT$ a sequential filling is observed. This filling forms the monolayer. When $\varepsilon/k_BT < w/k_BT$ the situation is just the opposite, two broad plateaus are formed previous to saturation. The first one (marked with an arrow in figure 2) corresponds to the filling of the vertices and edges of the bottom plane. The inset shows the partial coverage of each plane, for $\varepsilon/k_BT$ =-7.0. The isotherm of the bottom plane ($\theta_1$) shows a plateau that corresponds with the first plateau in the total isotherm. It is interesting to observe that $\theta_2$ and $\theta_3$ increase at the same chemical potential, which corresponds with the filling of the vertices sites on each plane. A green ellipse marks the mentioned phenomenon.

When the bottom plane is complete, ($\theta_1 \approx 0.33$) the other planes begin to fill until they saturate at $\theta_2=\theta_3 \approx 0.04$. This means the walls are decorated. At this point, all the surfaces exposed are completely decorated with the adsorbate. This corresponds with a second plateau of θ that includes the formation of the monolayer in the solid. The system minimizes its energy when it maximizes the interactions intra-particle. When the entire surface is decorated the only possible links are "inter-particles", then a condensation is observed. This behavior is typical of attractive interactions. For clarity, figure 3 shows the snapshot of the solid. Figure 3(a) represents the first plateau of $\varepsilon/k_BT<-5.0$ where a "square ring" on the bottom plane is observed. Figure 3(b) shows the state of occupation of the systems at the second plateau in the total isotherm.

Figure 4 shows the energy per site. As $\varepsilon/k_BT$ tends to more negative values, changes at slope are observed. This occurs at the coverage where the plateaus are formed in the adsorption isotherms. The black lines indicate this change at the first plateau. Other interesting thermodynamic parameter is the differential heat defined in ec. (4). Steps can be identified at coverage where the plateaus are present. The black arrow indicates the first step for $\varepsilon/k_BT = -7.0$.

One of the objectives of this work is to analyze the effect in the thermodynamics adsorption of the size $L$. Figure 5 shows the isotherms to $L=2, 5, 10, 20, 30$ for $\varepsilon/k_BT = -5.0$. For $L>5$ all the plateaus previously described are observed. The saturation coverage depends on $L$. As $L$ tends to zero, the walls begin to correlate, then no plateaus is observed. A question that we want to answer is whether the plateaus are observable for large values of L. To this purpose, the saturation coverage $\theta_{sat-A}$, the first and the second plateau $\theta_{1-A}$ and $\theta_{2-A}$ are written as function of $L$:

$$\theta_{1-A}(L) = \frac{4L}{V_{NH}} = \frac{4L}{3L^2} = \frac{4}{3L} \tag{6}$$

$$\theta_{2-A}(L) = \frac{N_{SNH}}{V_{NH}} + \frac{N_{surf}}{V_{NH}} = \frac{12L+L^2}{3L^2} + \left.\frac{10L+30l}{3L^2}\right|_{l=5} = \frac{L^2+22L+150}{3L^2} \tag{7}$$

$$\theta_{sat-A}(L) = \frac{3L^2+(L+l)^2}{V_{NH}} = \left.\frac{3L^2+(L+l)^2}{3L^2}\right|_{l=5} = \frac{4L^2+20L+100}{3L^2} \tag{8}$$

where, $N_{SMH}$ are sites on the walls and $N_{surf}$ are the sites on the external surface. These expressions are plotted versus $L$ in the inset of figure 5. The dots correspond to the MC simulations. A good concordance is observed. If $L \to \infty$, the walls are very far from each other and we get $\theta_{sat-A}(L) \to 4/3$, $\theta_{1-A}(L) \to 0$ and $\theta_{2-A}(L) \to 1/3$. The first plateau disappears as expected while the others reach a constant value. For the second plateau, in this approach, the area of the bottom plane is imposed on the other areas involved.

**3.2 Model B**

As explained before, the NH with a "curvature" is modeled in this section. Figure 6 shows the adsorption isotherms for several values of energies at $L=30$. As in the previous model, the isotherms saturate at different values, due to the fact that there are vacancies in the NH. However, as $\varepsilon/k_BT << w/k_BT$, such behavior disappears. When the energies change to more negative values, two broad plateaus are observed. The inset shows partial coverages, for $\varepsilon/k_BT =-10.0$. These coverages vary together, which is marked with a green circle. The walls are simultaneously filled in concordance with the formation of the first plateau in the total isotherm. The NH steps increase the intra-particles links. When $\theta_1$ saturates, it implies that the bottom plane is filled. The second plateau corresponds to the filling of the surface sites, including the monolayer. At this point, all the exposed surfaces are decorated with β-particles. After this, the NH is abruptly filled. A jump is observed in the isotherms until the saturation coverage. The snapshots are shown in figure 7 where (a) corresponds to the first plateau and b) to the second.

The behavior of the isotherms is shown in figure 8 for $\varepsilon/k_BT=-10.0$, and different values of $L$. It is important to mention that $L=5$ is the minimum size that conserves the characteristic of the NH. An incipient jump can be observed at very low coverage. The plateau is visible in the regimen $L\to 0$. All this behavior corresponds to the filling of corner sites of each plane. In these sites each adsorbate is linked to three particles of the solid. The inset (i) shows a zoom of this plateau. As before, recurrence relations for $L$ can be obtained.

$$\theta_{1-B}(L) = \frac{12L-24}{3L^2 - 12L + 20}, \tag{9}$$

$$\theta_{2-B}(L) = \frac{L^2 + 10L + 158}{3L^2 - 12L + 20}. \tag{10}$$

$$\theta_{sat-B}(L) = \frac{L^2 + 10L + 150}{3L^2 - 12L + 20} + 1. \tag{11}$$

where $\theta_{1-B}$, $\theta_{2-B}$ and $\theta_{sat-B}$ correspond to the first plateau, second plateau, and the saturation coverage respectively. These parameters are shown in the inset (ii). The points are obtained from simulations, and, as before, there is a good correspondence with the recurrence relations. For the approach $L\to\infty$, we obtain $\theta_{1-B}(L)\to 0$, $\theta_{2-B}(L)\to 1/3$ and $\theta_{sat-B}(L)\to 4/3$. The energy per site and the differential heat are in complete concordance with the isotherms. The energy presents a change of slope, and the differential heat shows the formation of a step for each regimen of deposition. As in model A, the first plateau disappears because the walls are very far from each other.

### 3.3 Model C

To consider explicitly the curvature effect, we can take into account a nanosystem with negative curvature. This model includes a NH in which its sites strongly correlate. Following these ideas we use the inner sites of half nanoparticle(NP) to define a NH. In our case we consider icosahedra NP where the depositions sites are $NP_{Bulk}$ = 50, 98 and 130 sites. Two sites are nearest neighbors if the distance between them is less than 3.5 Å. The energy is the same as before. The energy between the shell-bulk sites is $\varepsilon/k_BT$, and between bulk-bulk sites, $w/k_BT$. The monolayer is not considered. Figure 9 represents the isotherms normalized to $NP_{Bulk}$ =130. For $\varepsilon/k_BT$=-1.0, we observe a jump in the isotherms that can be related to a condensation, which means the NH is filled.

The decoration for $\varepsilon/k_BT$<-1.0 begins from the more linked sites, that is, the vertice sites. An incipient plateau is observed at low coverage, which is indicated with a black arrow in the isotherms. Subsequently, the sites on the walls are occupied. A broad plateau is formed around θ≈0.58. From the last plateau mentioned, a jump is observed until a third plateau at θ≈0.97. The remaining empty sites correspond to the most superficial sites. These sites are less connected because they are on the external border, in an opposite situation to the first plateaus reported. The inset (i) shows a zoom of this plateau. As in the cases previously analyzed, energy per site reports a change of slope at the same coverage as the plateaus. In the inset (ii) the differential heat $q_d$ is shown. Steps are indicated with arrows. The intensity of the steps depends on the values of $\varepsilon/k_BT$. Snapshots of a cross section for the second plateau and the third plateau are shown in figure 10. The black dots represent the atom substrate, the red ones adsorbates, and the white dots empty sites; the blue lines are guide eyes for the geometry. When the size of the NH is decreased, all the previous plateaus are shifted as can be observed. Figure 11 shows the dependence of isotherms with the different quantity of inner sites, for $\varepsilon/k_BT$=-5.0 and $w/k_BT$=-1.0. The inset shows the differential heat where the mentioned shift is observed.

**Conclusions**

In this paper we consider the deposition on nanoholes with three different lattice gas-models. Two kinds of energetic interactions have been considered: interparticles *w*, and intraparticle ε. Model A emulates a solid idealized with nanohole with vertical walls. Attractive interactions in all cases were considered since they can be associated to interaction between metals. The thermodynamics analysis has been done in function of the variation of $\varepsilon/k_BT$ and several sizes of NH. The isotherms show the formation of two plateaus. These are related to the different regimens of decoration in the solid. The first plateau is associated to deposition in vertices and edges of the bottom plane. The second plateau is associated with the formation of the monolayer and the decoration of all the walls. The isosteric heat and the energy per site are in concordance with the isotherms results. When the size of the NH is changed, the isotherms have a different regimen ofsaturation, this can be recognized as nanoeffect. The recurrence equations allow analyze the tendencies for high values of L, and the first plateau disappears at this limit (the border effect is negligible), while the second reaches a constant value. At this limit it is reasonable to think that at very large distances between the walls the system can be considered a two-dimensional substrate. Another important conclusion is that the formation of the monolayer occurs before the filling of the nanohole.

The B model includes a curvature on the walls. In this case there are three stages of deposition: the first corresponds to the adsorption on the vertex sites, this nanoeffect is only visible at very low values of *L*. The second corresponds only to the filling of the walls, and finally the last plateau corresponds to the filling of the bottom plane and the monolayer. Both models have intermediate states before the full filling of the NH. The previous states depend on the distribution of lateral links in the model. In model B, there is no preference between facets or bottom sites. This general behavior does not depend on the quantity of the planes considered. Even if geometry (111) is considered, the general behaviors do not change.

Model C considers a NH with the internal geometry of the icosahedra, which presents a distribution of links that depends on the position in the NH. This model confirms the tendencies to cover the walls firstly.

As a general conclusion it can be stated that the models exhibit a decoration of the walls before the full filling of the NH. The filling of the NH is done through an abrupt condensation. From the recurrence relations we can affirm that these behaviors are consistent for all values of the size L.

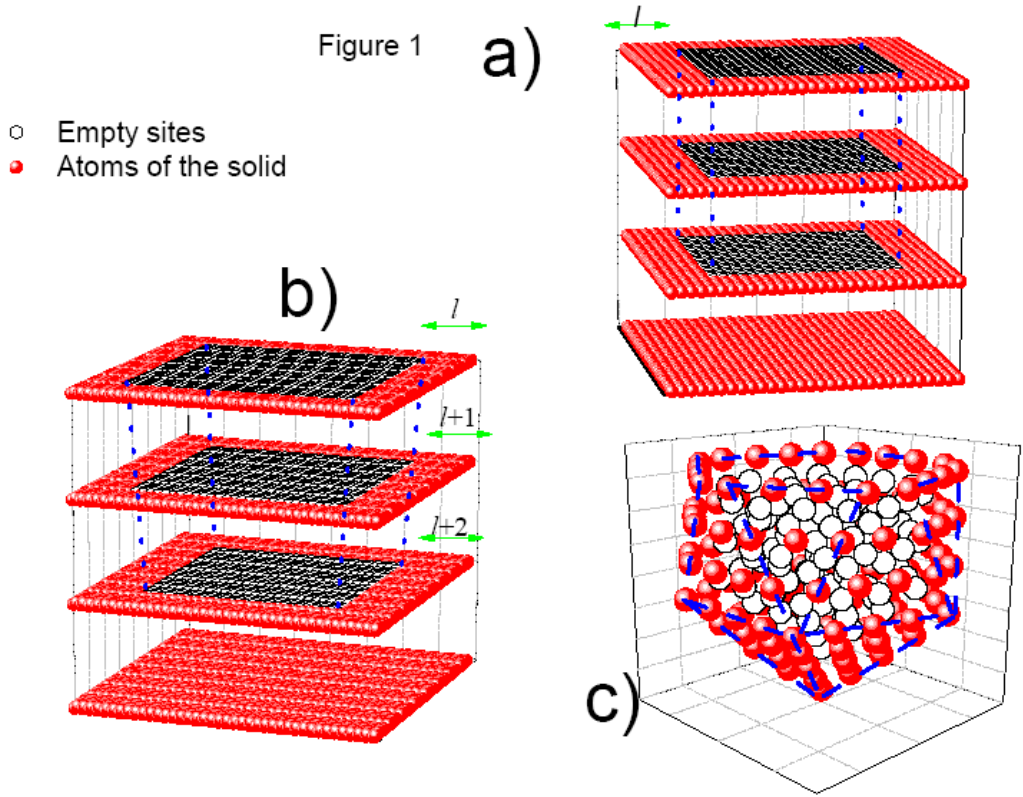

Figure 1

○ Empty sites
● Atoms of the solid

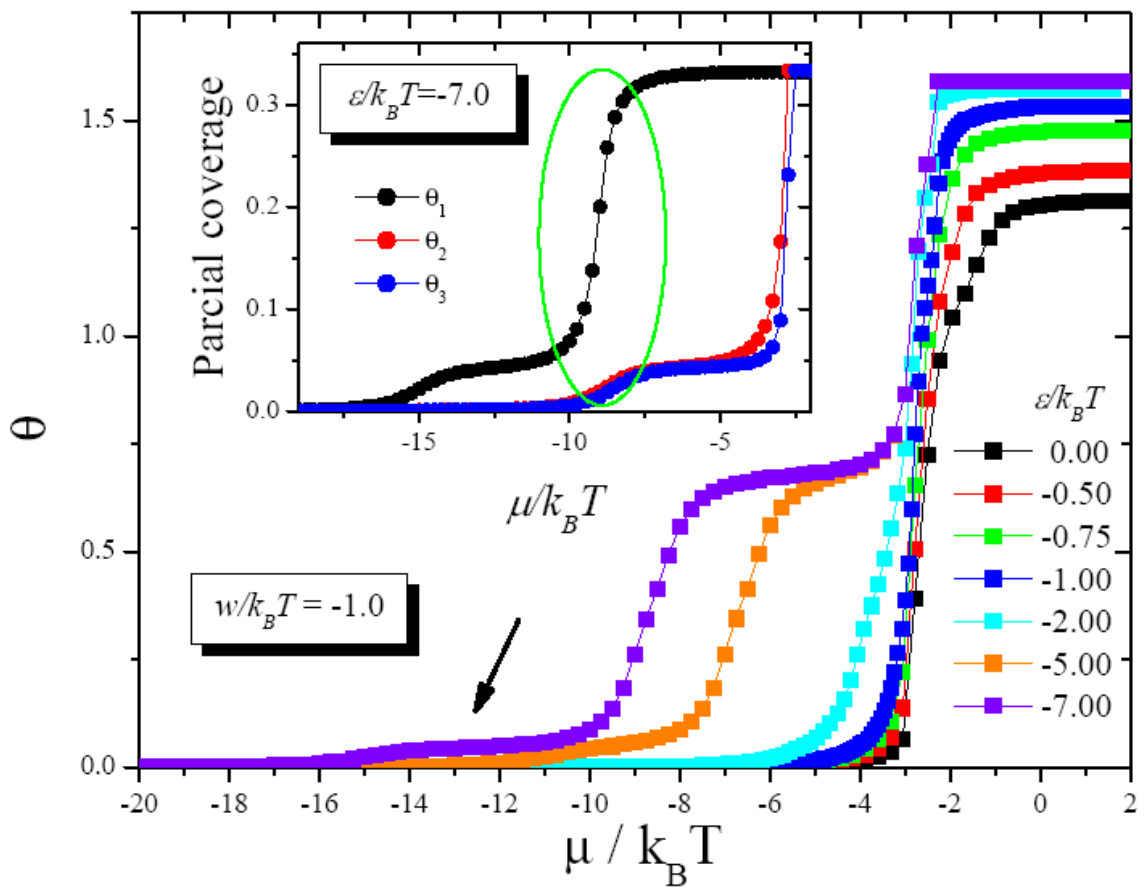

figure 2

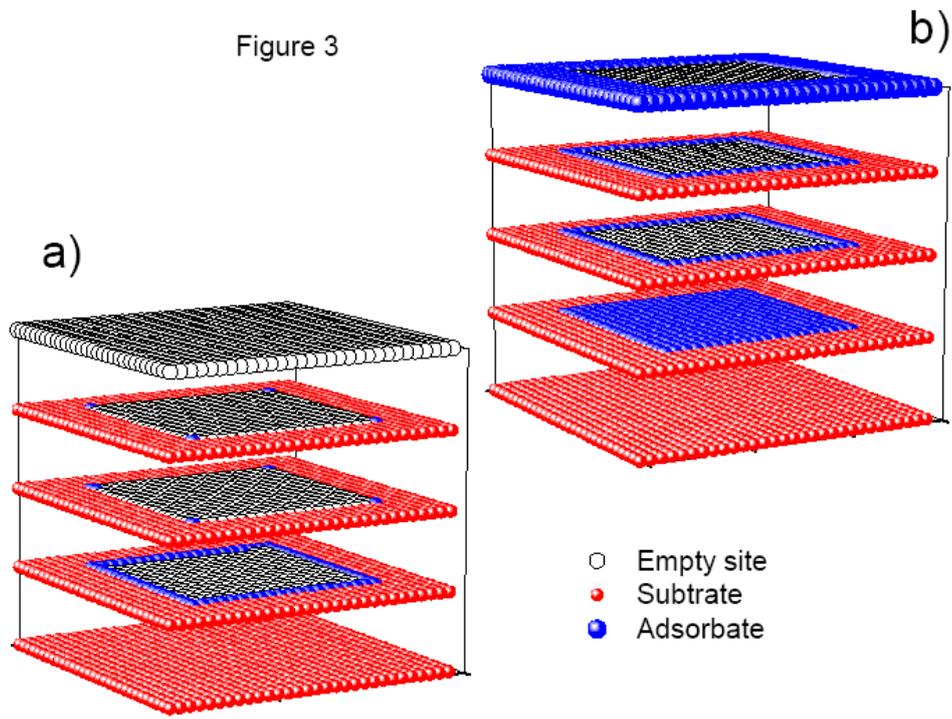

Figure 3

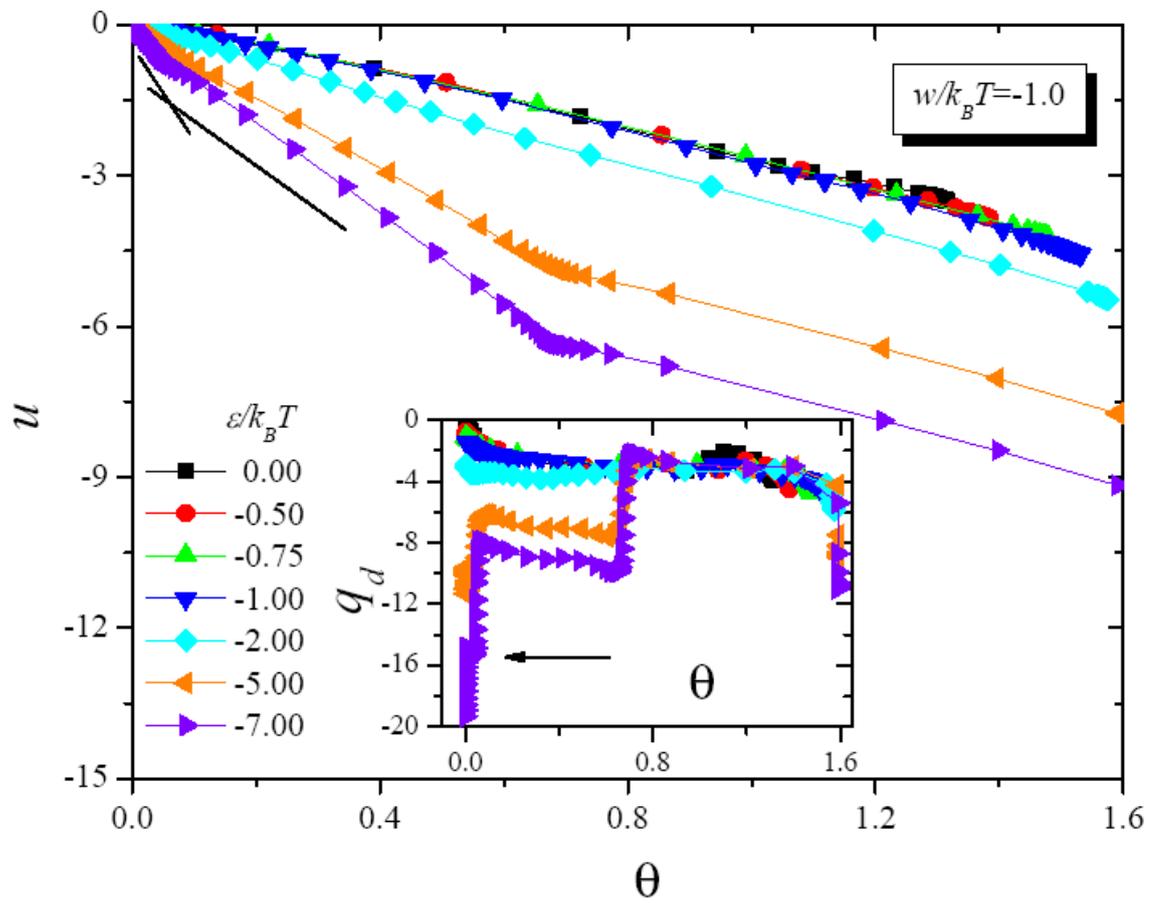

figure 4

figure 5

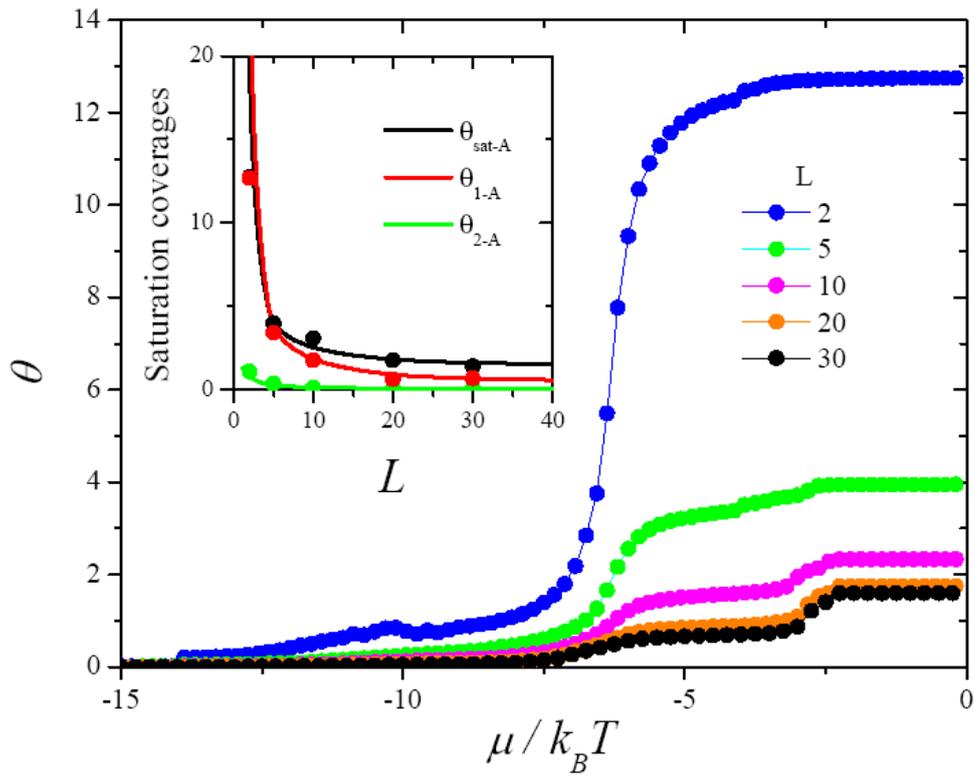

Figure 6

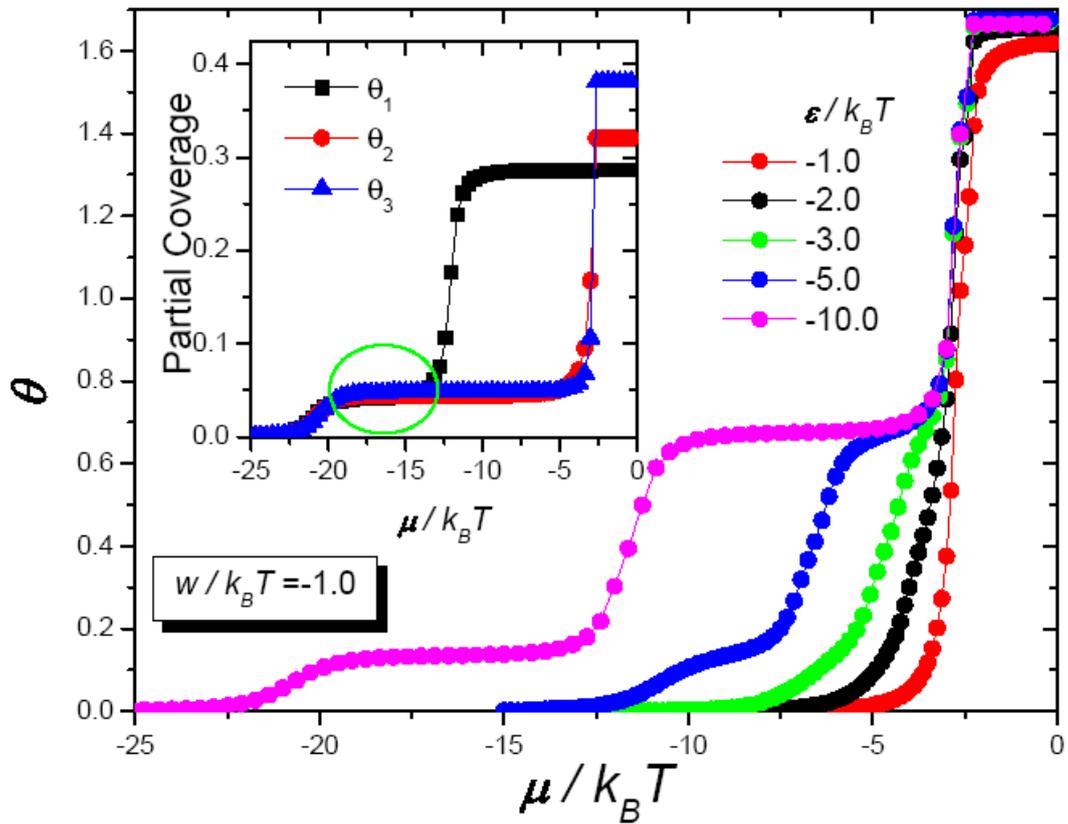

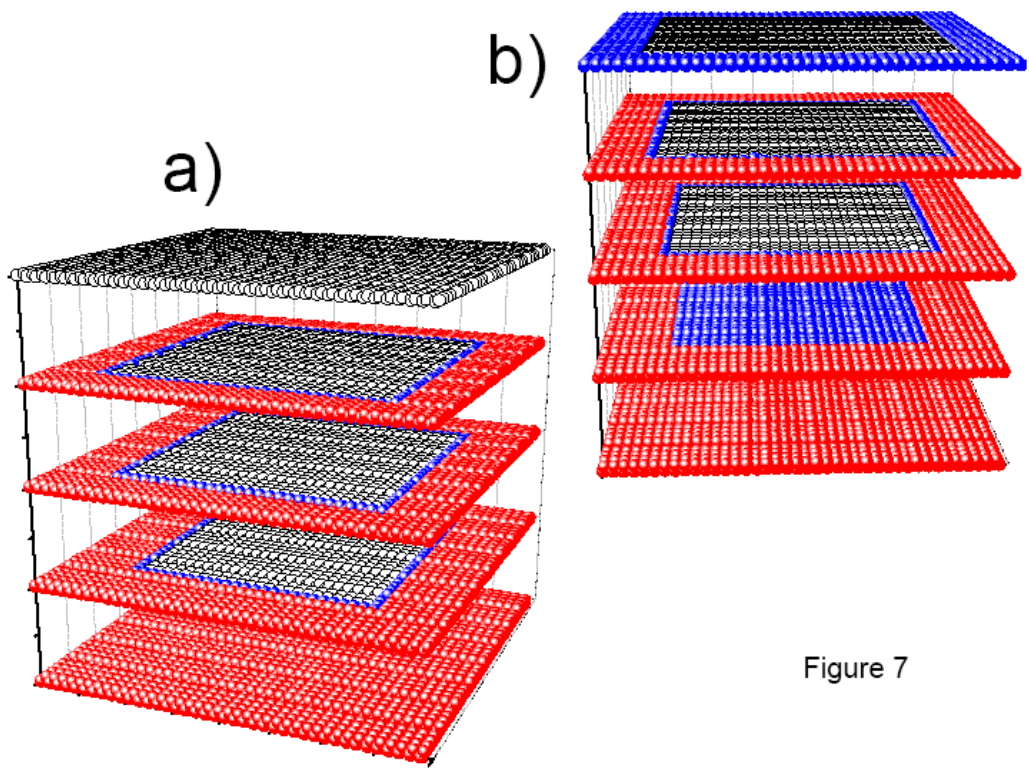

Figure 7

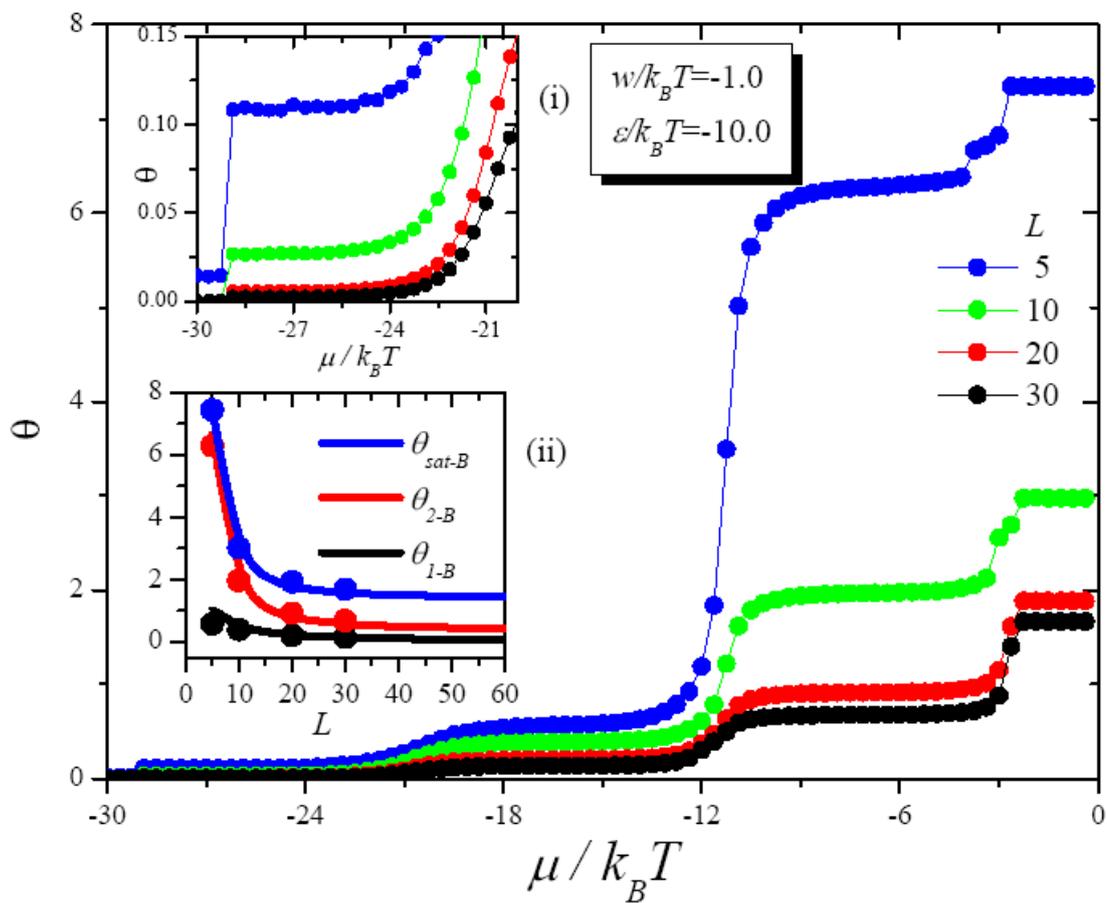

figure 8

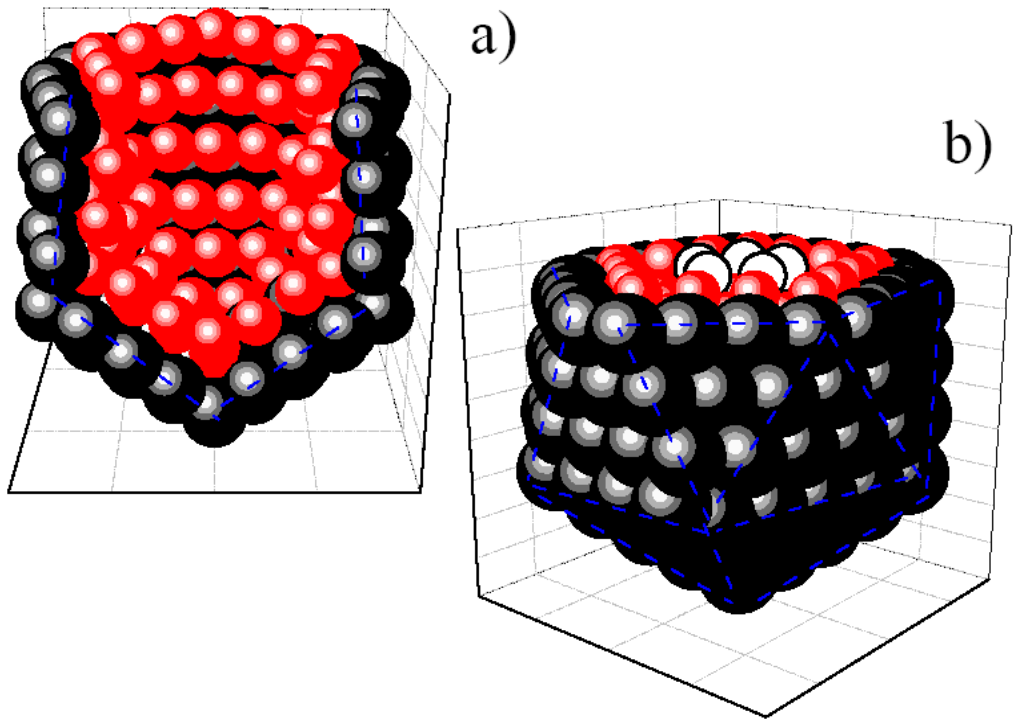

Figure 10

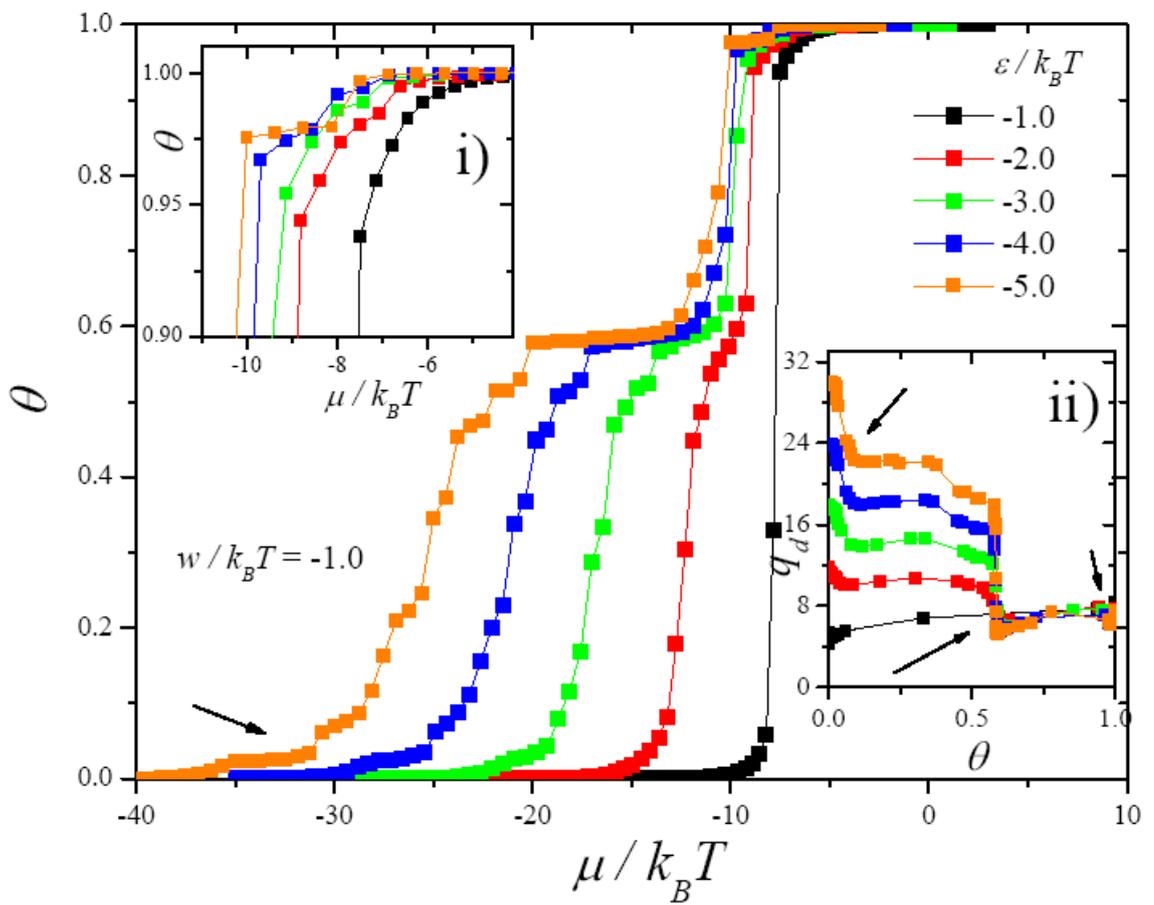

Figure 9

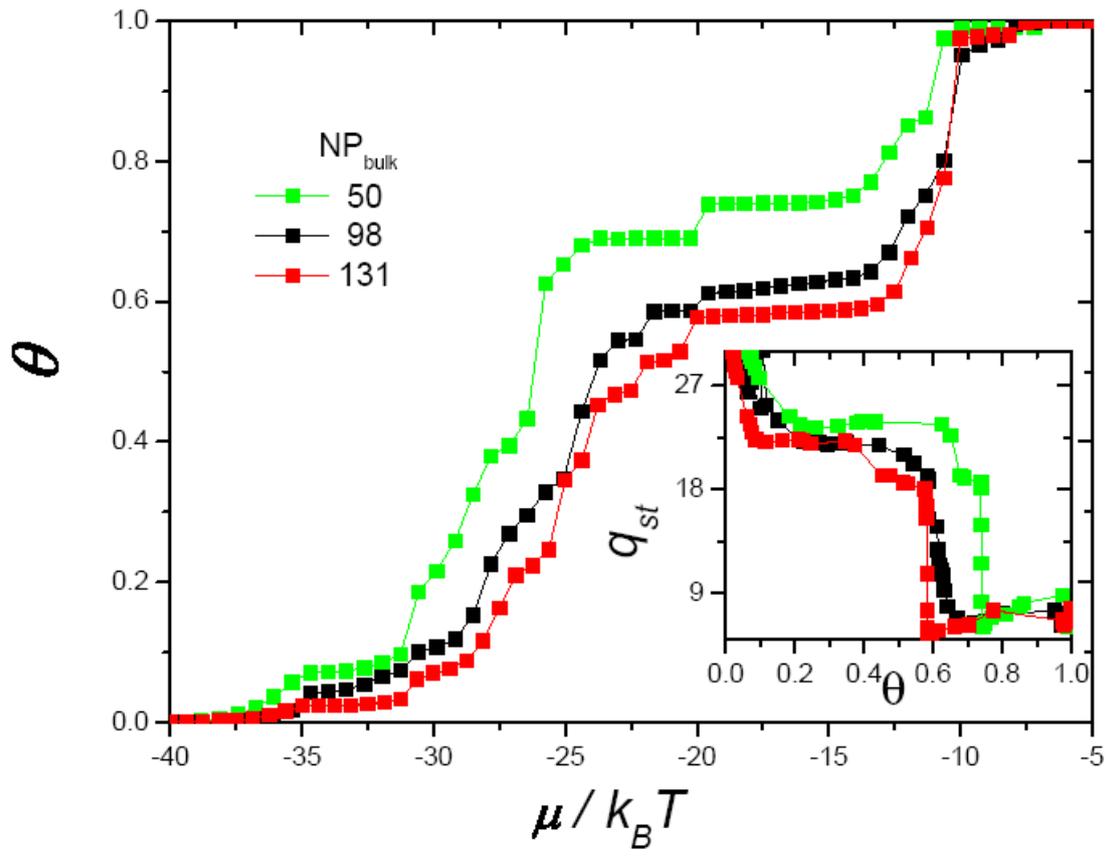

Figure 11